\documentclass{aa}
\usepackage{psfig}

\def\bron{GRS~1747-312}
\def\ecs{erg~cm$^{-2}$s$^{-1}$}
\def\lum{erg~s$^{-1}$}

\begin{document}
\thesaurus{05(08.09.2 \bron; 08.14.1; 10.07.3 Terzan~6; 13.25.1)}

\title{The orbital period of the recurrent X-ray transient in Terzan~6}
\author{J.J.M.~in~'t~Zand\inst{1}
 \and A.~Bazzano\inst{2}
 \and M.~Cocchi\inst{2} 
 \and R.~Cornelisse\inst{1,3}
 \and J.~Heise\inst{1}
 \and L.~Kuiper\inst{1}
 \and E.~Kuulkers\inst{1,3}
 \and C.B.~Markwardt\inst{4}
 \and J.M.~Muller\inst{1,5} 
 \and L.~Natalucci\inst{2}
 \and M.J.S.~Smith\inst{1,6}
 \and T.E.~Strohmayer\inst{4}
 \and P.~Ubertini\inst{2} 
 \and F.~Verbunt\inst{3}
}
\offprints{J.J.M.~in~'t Zand (at e-mail {\tt jeanz@sron.nl})}

\institute{     Space Research Organization Netherlands, Sorbonnelaan 2,
                NL - 3584 CA Utrecht, the Netherlands
         \and
                Istituto di Astrofisica Spaziale (CNR), Area Ricerca Roma Tor
                Vergata, Via del Fosso del Cavaliere, I - 00133 Roma, Italy
	 \and
		Astronomical Institute, Utrecht University, P.O. Box 80000,
		NL - 3508 TA Utrecht, the Netherlands
	 \and
	 	NASA Goddard Space Flight Center, Code 662, Greenbelt, MD 20771, 
		U.S.A.
         \and
                BeppoSAX Science Data Center, Nuova Telespazio,
                Via Corcolle 19, I - 00131 Roma, Italy
         \and
                BeppoSAX Science Operation Center, Nuova Telespazio,
                Via Corcolle 19, I - 00131 Roma, Italy
                        }
\date{Received, accepted }
\maketitle

\begin{abstract}
Four or five new outbursts were detected of the bright X-ray transient \bron\ 
in the 
globular cluster Terzan~6 between 1996 and 1999, through monitoring campaigns 
with the Wide Field Cameras (WFCs) on {\em BeppoSAX\/} and the Proportional 
Counter Array (PCA) on {\em RossiXTE\/}. This is the first time that the
source is seen to exhibit recurrent outbursts after the discovery
in September 1990 with ART-P on {\em Granat}. Three target-of-opportunity 
observations in 1998 and 1999, with the 
narrow-field instruments on {\em BeppoSAX\/} and the PCA,
revealed one sharp drop in the flux which we
identify as an eclipse of the compact X-ray source by
the companion star. A detailed analysis of WFC data identifies
further eclipses and we measure the orbital period
at 12.360$\pm$0.009~hr. This is consistent with an identification as a 
low-mass X-ray binary, as suggested already by the association with a globular
cluster. The eclipse duration is $0.72\pm0.06$~hr. This implies that the
inclination angle is larger than 74\degr.
The 0.1-200~keV unabsorbed peak luminosity is $7\times10^{36}$~\lum. 
The nature of the compact object is unclear.
\end{abstract}

\section{Introduction}
\label{intro}

Currently, 12 bright X-ray sources are known in 12 different globular clusters.
Presumably they are all low-mass X-ray binaries (LMXBs) but orbital
periods have so far been determined for only four of them (NGC~6441,
NGC~6624, NGC~6712 and NGC~7078). These four periods range between
0.19~hr (for NGC~6624, Stella et al. 1987) and 17.10~hr (for NGC~7078, 
Ilovaisky et al. 1993).
An interesting characteristic of these is that a high percentage 
(11 out of 12) exhibit type-I X-ray bursts which are thought to originate 
on the surfaces of neutron stars. For the
bright LMXBs in the Galactic plane this is only $\sim$30\%. 
This paper deals with the sole bright X-ray source in a 
globular cluster that has so far not exhibited bursts, namely \bron\
in Terzan 6.

Terzan 6 (Terzan 1968) is a metal-rich globular cluster at a distance of
7~kpc and highly reddened, with E(B-V)=2.24 (Barbuy et al. 1997). The 
cluster has undergone core collapse, its
core and tidal radii are 3.3\arcsec\ and 316\arcsec\, respectively
(Trager et al. 1995). 

The transient X-ray source \bron\ was first detected in early September, 1990, 
both with the X-ray telescope ART-P on {\em Granat\/} (Pavlinsky et al. 1994)
and with ROSAT (Predehl et al. 1991).
The ART-P spectrum was compatible with 5.8~keV thermal bremsstrahlung and
absorption by cold gas of cosmic abundances with 
$N_{\rm H}=6\times10^{22}$~cm$^{-2}$.
The flux was $5.2\times10^{-10}$~\ecs\ between 4 and 12 keV or 0.03 Crab
flux units between 2 and 10 keV. The ROSAT all-sky survey measurement 
reveals the
most accurate X-ray position so far: 
R.A.~=~17$^{\rm h}$50$^{\rm m}$46.6$^{\rm s}$, 
Decl.~=~-31$^\circ$16\arcmin40\arcsec\ (Eq. 2000.0, 1$\sigma$ 
error radius 20\arcsec, Verbunt et al. 1995). This is 0.9 
core radii from the center of Terzan 6 as published
by Barbuy et al. (1997).
Furthermore, this ROSAT measurement is compatible with 
the hard bremsstrahlung spectrum of 5.8~keV, but the
absorption column is rather lower at $N_{\rm H}=2\times10^{22}$~cm$^{-2}$ 
(Verbunt et al. 1995). We have re-analyzed this spectrum, and
find that a higher column, more in line with what was found
from ART-P data, is not acceptable.

No other outbursts by \bron\ have been
reported. However, Terzan 6 was not observed with either the sensitive 
X-ray instruments on {\em Einstein\/} or
EXOSAT. Rappaport et al. (1994) observed Terzan 6 with the HRI on ROSAT 
on March 16-17, 1992, and 
found no source within 10 core radii. Thus, they confirmed that
the source is a transient. Their HRI upper limit extrapolates
to a PSPC upper limit which is a factor of 150 lower than the PSPC detection 
in September 1990 (Verbunt et al. 1995).

In this paper five new X-ray data sets are discussed which were
obtained since 1996. The nature of the data
is twofold. Monitoring observations were obtained with the Wide Field 
Cameras (WFCs) on {\em BeppoSAX\/} between August 1996 and April 1999, and
with the Proportional Counter Array (PCA) on {\em RossiXTE} since February
1999. Sensitive target-of-opportunity observations (TOOs) were 
carried out 
with the narrow-field instruments (NFI) on {\em BeppoSAX\/} in September
1998 and with the PCA in September 1998 and in June 1999.
We present an analysis of all these observations and speculate on the nature 
of the source.

\begin{figure}[t]
\psfig{figure=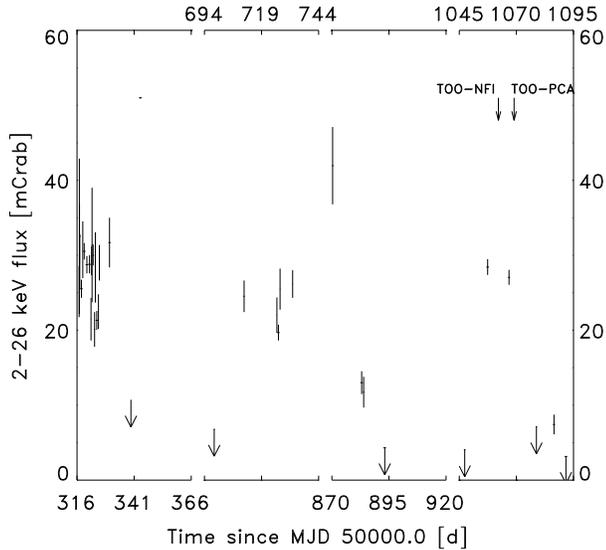,width=\columnwidth,clip=t}

\caption[]{Light curve as measured with the WFCs in the 2 to 26 keV band.
The time resolution is 1 {\em BeppoSAX\/} observation period (i.e.,
typically 1 day). The intensity is
corrected for dead time and background. Only data points within 20 days
from a detection have been considered. The large arrows indicate 3$\sigma$
upper limits
\label{figwfclc}}
\end{figure}

\begin{figure}[t]
\psfig{figure=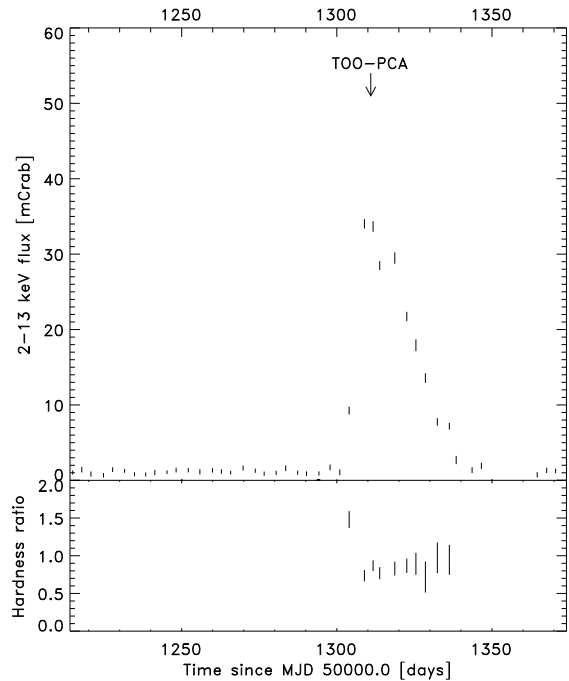,width=\columnwidth,clip=t}

\caption[]{Light curve as measured with the PCA in 2 to 13 keV (upper panel)
and the hardness ratio which is defined as the ratio of the photon count rate 
in the 6 to 30 keV band to that that in the 2 to 6 keV band (lower panel).
The integration time per data point is about 60~s.
\label{figpcalc}}
\end{figure}

\section{Monitoring observations with the WFCs}
\label{secwfcobs}

The WFCs (Jager et al. 1997) on the {\em BeppoSAX\/}
satellite (Boella et al. 1997a) carry out a long-term program of monitoring
observations of the 40$^{\rm o}\times40^{\rm o}$ field around the Galactic
center. One of the goals is to detect X-ray 
transient activity from LMXBs whose
Galactic population exhibits a strong concentration in this field (e.g.,
Heise 1998 and Ubertini et al. 1999). With respect to the 
monitoring observations with the All-Sky Monitor on {\em RossiXTE\/}
the WFC monitoring observations are not as much source confused so close
to the Galactic center and are five times as sensitive for 1-day time spans.

The program consists of campaigns during the spring and fall of each year. 
Each campaign lasts about two months and typically comprises weekly
observations. Up to mid 1999, a total net exposure
time of 2.6 million seconds was accumulated during 6 campaigns.

Terzan 6 is only 2\fdg6 distant from the direction of the Galactic Center 
which is the pointing direction of the monitoring
program. Therefore, Terzan 6 is observed near to the optimum sensitivity. 
Fig.~\ref{figwfclc} presents the light curve. At least three outbursts 
are recognized with first detections at 
MJD~50316, 50697 and 51047. The sampling is far from complete.
There is a gap between MJD~50732 and MJD~50869 during which two dates the 
source was active. If the source returned to quiescence in this
gap, which seems likely given the durations of other outbursts, we have 
detected four outbursts. The incomplete sampling
makes it difficult as well to assess the peak flux of each outburst.
The maximum intensity over all accurate detections is about 30 mCrab in 
the 2 to 26 keV
band. This happens to be quite close to the flux measured with ART-P in the 
1990 outburst (Pavlinsky et al. 1994).

For each of the seven observations where \bron\ was detected above a 
signal-to-noise ratio of 20 we determined the position. The average
of these is R.A.~=~17h~50m~46.3s, Decl.~=~--31\degr~16\arcmin~42\arcsec\
(Eq. 2000.0) and the 99\% confidence level error radius 0\farcm7.
This position is only 4\arcsec\ (nominally) from the ROSAT position of the 
source detected in 1990 and on top of Terzan~6.

\begin{figure*}[t]
\psfig{figure=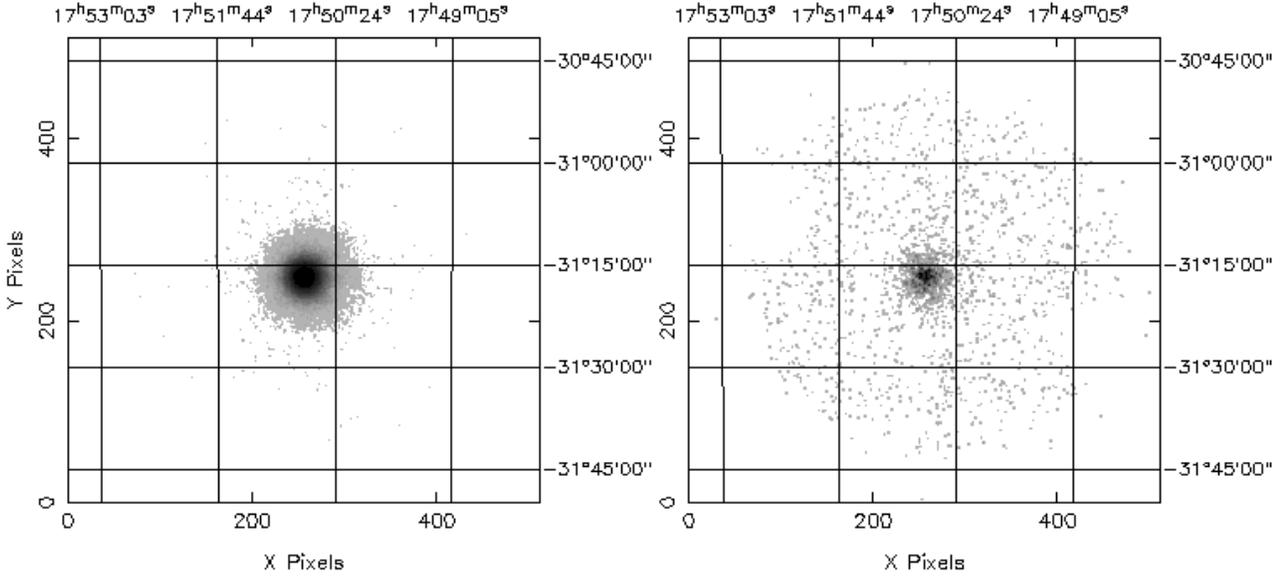,width=2\columnwidth,clip=t}

\caption[]{Image of the MECS outside the intensity drop (left, from 0 to 23.4 
ks after MJD~51062.2618 and from 32.4~ks to observation end) and during the 
drop (right, between 26.2 and 28.4~ks after MJD~51062.2618). 
There is clearly residual emission during the drop. Both these images
have as pixel unit counts and were smoothed through a convolution with 
a Gaussian of standard
deviation 1 pixel in both X and Y. The scaling is between 0 (white) and the
respective maximum levels (black, 388 counts left and 20 counts right).
\label{figplaat}}
\end{figure*}

\section{Monitoring observations with the PCA}
\label{secpcabulgeobs}

Since February 1999, scanning observations are carried
out with the PCA on {\em RossiXTE\/} of a 
rectangular region surrounding the Galactic center, of approximately
16\degr$\times$18\degr\ in size and on a semi-weekly basis. In addition 
to the known persistently bright X-ray sources, of which about 20 are 
detected, old and new transient sources are found. 
The sensitivity of these scan observations is about 1~mCrab over the whole
region. This is one to two orders of magnitude more sensitive than the 
observations with the All-Sky Monitor on 
{\em RossiXTE\/}. Thanks to the high frequency of the observations and
the high duty cycle throughout the year ($>80$\%), it is possible to uniformly 
sample outburst light curves of relatively faint transients. In contrast,
the WFC observations have a lower duty cycle of $\sim30$\% and an observation
frequency at least twice as low. Terzan 6 is
covered by all scan observations. It was found to go into outburst starting 
May 5, 1999, at MJD~51303.97. The light curve is presented in 
Fig.~\ref{figpcalc} (upper panel). 
The observations appear to sample the light curve well, there
is no strong variability from data point to data point. Therefore, the 
peak flux can be confidently measured. The profile is a fast-rise 
exponential-decay function which is so typical of many other LMXB transients. 
The rise time was observed to be shorter than 1 week
and could well have been much shorter.
The peak intensity is 350.1$\pm$6.5~cts~s$^{-1}$ per 5~proportional
counter units and occurred on 
MJD~51308.9, or 34.0$\pm$1.1~mCrab (this includes a 
systematic uncertainty of 0.9~mCrab for the quiescent level which may
be contaminated by diffuse Galactic emission). 
The outburst intensity dropped
to below the detection limit somewhere between MJD~51338 and 51343. The 
activity was seen for a 35 day period. The
e-folding decay time is $18.1\pm0.5$~d, if one fits an exponential function
to all data beyond and including the time of peak intensity.
Until mid 1999 this is the only 
outburst of \bron\ seen in about 160 days of PCA scan observations. 
The other outbursts (i.e., measured with WFC and ART-P) have not shown 
fluxes that are significantly higher than the peak flux measured in this 
outburst.

The decay time of 18~d is reasonably consistent with the rate of decay
of the latter two WFC-detected outbursts. The situation is unclear
for the former two outbursts due to incomplete sampling. 

The evolution of the 6--30 to 2--6 keV hardness ratio is also presented in 
Fig.~\ref{figpcalc} (lower panel). This shows that the spectrum is
significantly harder during the rise than during the decay, and that
there is no measurable spectral change during the decay (there 
is only a hint of a slight hardening during decay).

\section{Broad-band spectral measurements with {\em BeppoSAX\/}-NFI}
\label{secspec}

\subsection{Observation and data reduction}

The NFI include 2 imaging instruments 
that are sensitive at photon energies below 10~keV: the Low-Energy and the 
Medium-Energy Concentrator Spectrometer (LECS and MECS, see 
Parmar et al. 1997 and Boella et al. 1997b, respectively). They have circular 
fields of view with diameters of 37\arcmin\ and 56\arcmin\ and effective 
bandpasses of 0.1-10.5 and 1.6-10.5~keV, respectively. The other two, non-imaging, 
NFI instruments are the Phoswich Detector System (PDS) which covers $\sim12$ 
to 300 keV (Frontera et al. 1997) and the High-Pressure Gas Scintillation 
Proportional Counter (HP-GSPC) which covers 4 to 120 keV (Manzo et al. 1997). 

A target-of-opportunity observation (TOO) was performed with the NFI on 
September 6.26-6.79, 1998. The net exposure times are 11.8~ks
for LECS, 22.9~ks for MECS, 10.5~ks for HP-GSPC and 10.4~ks for PDS. 
\bron\ was strongly detected in all instruments.
The LECS and MECS images show only one bright source (Fig.~\ref{figplaat}), 
the position as determined from the MECS image is consistent with that from 
WFC. We applied extraction radii of 8\arcmin\ and 4\arcmin\ for photons from 
LECS and MECS images, encircling at least 
$\sim95$\% of the power of the instrumental point spread function, to obtain 
light curves and
spectra. Long archival exposures on empty sky fields were used to define the
background in the same extraction regions. These are standard data sets
made available especially for the purpose of background determination. 
All spectra were rebinned so as to
sample the spectral full-width at half-maximum resolution by three bins
and to accumulate at least 20 photons per bin. The latter will ensure the
applicability of $\chi^2$ fitting procedures. A systematic error of 1\% was
added to each channel of the rebinned LECS and MECS spectra, to account
for residual systematic uncertainties in the detector calibrations 
(e.g., Guainazzi et al. 1998). The bandpasses were limited to 0.8--4.0 keV
(LECS), 2.2--10.5~keV (MECS), 4.0--25.0 keV (HP-GSPC) and 15--100 keV (PDS)
to avoid photon energies where either the spectral calibration of the 
instruments
is not complete or no flux was measured above the statistical noise. 
In spectral modeling, an allowance was made to
leave free the relative normalization of the spectra from LECS, PDS and HP-GSPC
to that of the MECS spectrum, to accommodate cross-calibration uncertainties
in this respect. Publicly available instrument response functions and software
were used (version November 1998).

For the non-imaging PDS and HP-GSPC, source confusion may be an issue,
particularly in the densely populated Galactic center field. We
verified whether any other point sources were active inside the field
of view (1\fdg1 FWHM for the HP GSPC and 1\fdg3 FWHM for the PDS). For these
instruments 5 pointings are relevant: one on-source pointing and two off-source
pointings per instrument. For our observation, the off-source pointings were
at +/--3\fdg5 and +/--3\fdg0 from the on-source pointings for the PDS and 
HP GSPC respectively (i.e., +/-- in the direction of the satellite positive 
and negative Y-axis). We used near-to-simultaneous WFC images to provide 
information on what sources were active inside the field of view. For 
the on-source pointing, the nearest active point source turned out to
be at a distance of 1\fdg4. This is outside the field of view of both 
instruments . For the two off-source pointings of the PDS, we find that 
the nearest active source is at 2\fdg6. For the off-source pointings of 
the HP-GSPC, the nearest active source is at 2\fdg5. We conclude that
source confusion is no issue in our observation.

\begin{figure}[t]
\psfig{figure=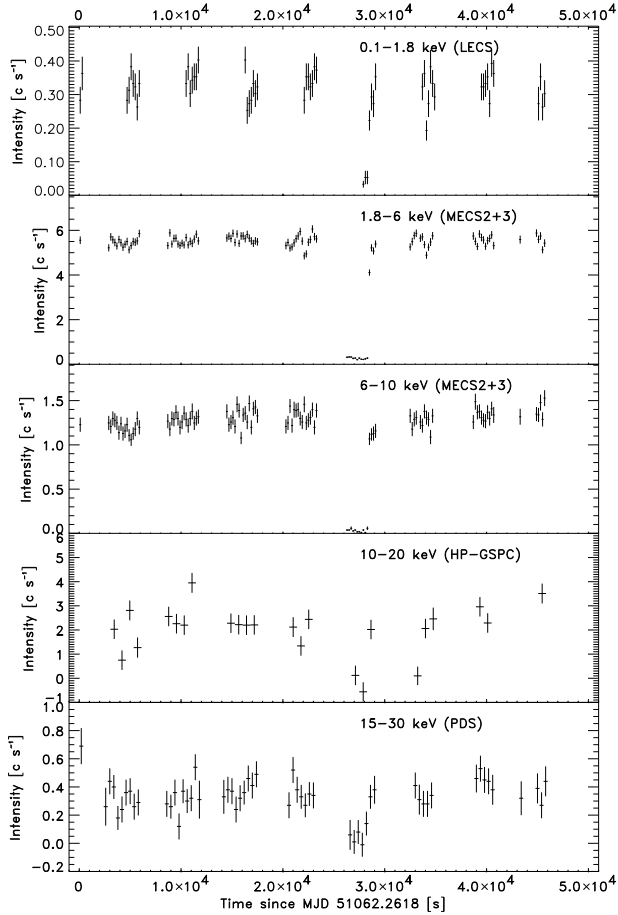,width=\columnwidth,clip=t}

\caption[]{Light curve as measured with the NFI in a number of bandpasses, 
corrected for background. The time resolution is 200~s except for that from 
the HP-GSPC which has a resolution of 764~s. The background 
levels are from top to bottom
$7.1\times10^{-3}$, $2.8\times10^{-3}$, $2.1\times10^{-3}$, 30.0 and
1.4~c~s$^{-1}$
\label{fignfilc}}
\end{figure}

\begin{figure}[t]
\psfig{figure=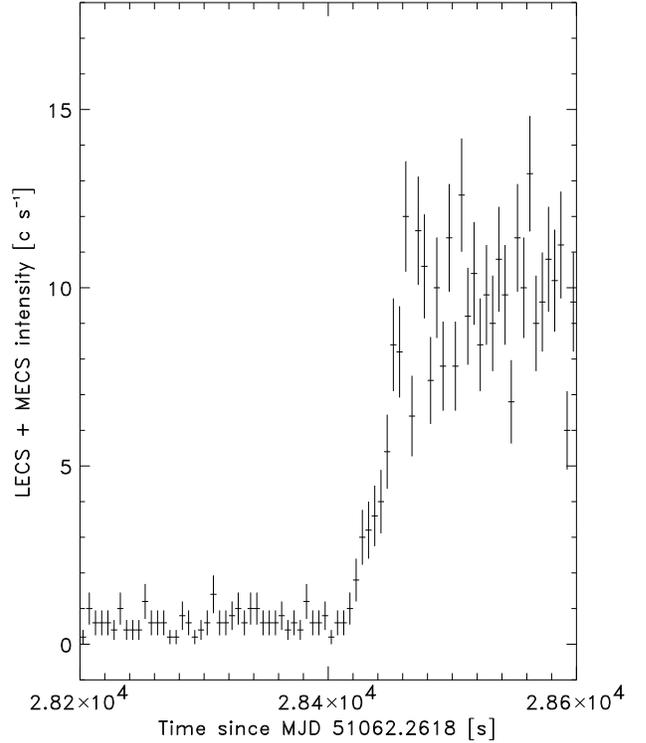,width=\columnwidth,clip=t}

\caption[]{Light curve zoomed-in at the time of emergence from the eclipse.
All LECS and MECS photons were added, including photons with energies
outside the bands defined in Fig.~\ref{fignfilc}, to increase the statistics.
No background subtraction was carried out. The time resolution is 5~s.
\label{figlczoom}}
\end{figure}

\begin{figure}[tb]
\psfig{figure=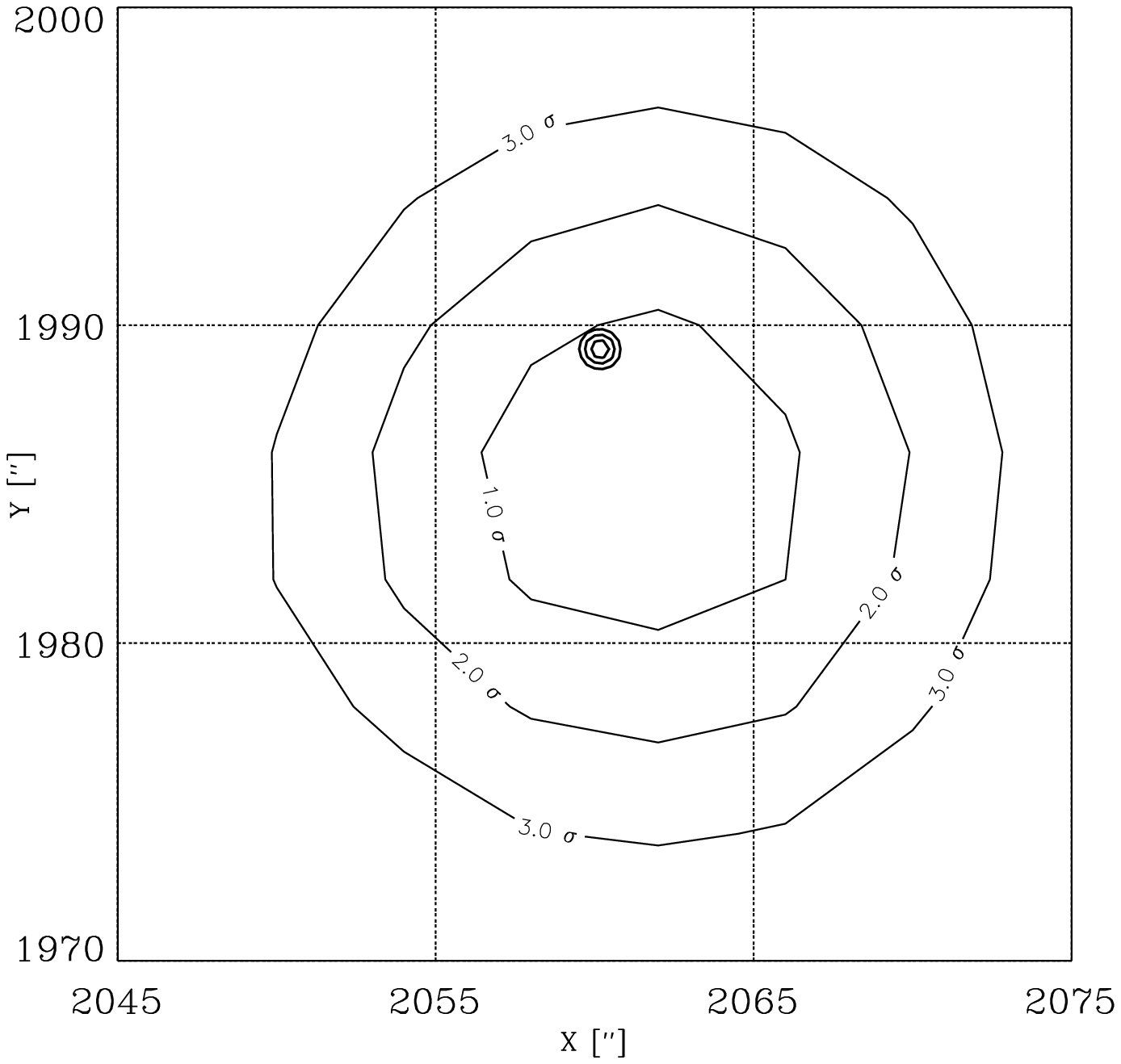,width=\columnwidth,clip=t}

\caption[]{Location confidence contours of point source during eclipse
(large contours) and outside eclipse (small contours). 
The image position is in instrument coordinates. It,
therefore, is not subject to the systematic errors that are
introduced by a transformation to celestial coordinates.
\label{fignfiml}}
\end{figure}

\subsection{Detection of an eclipse}
\label{secnfilc}

Fig.~\ref{fignfilc} shows the light curve in various bandpasses from the 
four NFI. Within the observation, the source 
shows a moderate intensity increase of about 10\% from beginning to end, and 
one deep 
drop starting between 23.4 and 26.2~ks after the start of the observation.
The drop lasts between 2.2 and 5.0~ks.
A zoomed-in light curve (Fig.~\ref{figlczoom}, including all LECS and MECS
data to increase the sensitivity) shows that the emergence from the drop
is not discrete but lasts about 35~s. For the remainder, the drop is flat 
for over 2200~s. This strongly suggests the drop to be caused by an eclipse 
by a companion star. 

The eclipse is not complete as evidenced by the MECS image 
accumulated during times of minimum intensity (see 
Fig.~\ref{figplaat}). To reliably establish that the faint source is
the same as the bright one and to accurately determine the intensity of
the faint source, we have carried out a maximum likelihood analysis of
the 2.2-10.5~keV MECS photons inside and outside the eclipse. 
In this analysis, point sources are searched for on top of a flat
background. It renders quantitative information on the source flux
and detection significance (e.g., Kuiper et al. 1998).
The image accumulation
times are 26.200 to 28.415~ks after the start of the observation for the 
eclipse and 0.0 to 23.400 plus 32.380 to 46.780~ks for outside the eclipse. 
In both images we found evidence for only a single point source.
Fig.~\ref{fignfiml} shows the location confidence contours 
of these sources. The best-fit positions are only 4\arcsec\ apart and
consistent with each other. Therefore, we reject the possibility that the 
source during the eclipse is other than outside the eclipse. The 2.2-10.5~keV 
count 
rate during the eclipse is $0.277\pm0.012$~c~s$^{-1}$. This is consistent with
what we found following the standard analysis (i.e., 
the rate then is $0.268\pm0.012$~c~s$^{-1}$ which
is $\sim$3\% lower than the aforementioned value as may be expected to be lost
in the
wings of the point spread function beyond the 4\arcmin\ accumulation radius).
 This,
furthermore, shows that the data analysis is not adversely affected
by diffuse emission (from, for example, the Galactic ridge).

The maximum likelihood analysis results in a position of 
R.A.~=~17$^{\rm h}$50$^{\rm m}$45.7$^{\rm s}$, 
Decl.~=~-31$^\circ$16\arcmin46.7\arcsec\ (Eq. 2000.0, 1$\sigma$ 
systematic error radius 50\arcsec). This position is 13\arcsec\ (nominally)
from the position of \bron\ (Verbunt et al. 1995).

\subsection{Spectrum}
\label{secnfispec}

\begin{table}[tb]
\caption[]{Parameter values of three model fits to the NFI spectrum
well outside the eclipse (i.e., at times before 23.4 and after 
32.4~ks after MJD~51062.2618). $\Gamma$ is the photon index. $N_{\rm H}$ 
is in units of 
10$^{22}$~cm$^{-2}$. The last line of each model specifies the 
$\chi^2_{\rm r}$ values for the fit without a bb (black body) 
component. These latter values apply
are after re-fitting the remaining parameters}
\begin{tabular}{ll}
\hline
Model              & high-energy cut off power law \\
                   & ($N(E)(:)E^{-\Gamma}$ for $E<E_{\rm cutoff}$ and\\
		   & $(:)E^{-\Gamma}{\rm exp}(-(E-E_{\rm cutoff})/E_{\rm fold})$\\
		   & for $E>E_{\rm cutoff}$ )\\
                   & + black body\\
$N_{\rm H}$        & $1.40\pm0.09$ \\
bb $kT$            & $2.0\pm0.1$ keV \\
bb $R/d_{\rm 7~kpc}$  & $1.5\pm0.2$ km \\
$\Gamma$           & $1.03\pm0.16$ \\
$E_{\rm cutoff}$   & $2.5\pm0.2$ keV\\
$E_{\rm fold}$     & $6.6\pm0.8$ keV\\
$\chi^2_{\rm r}$   & 1.27 (109 dof) \\
$\chi^2_{\rm r}$ without bb & 1.45 (111 dof)\\
\hline
Model                 & bremsstrahlung + black body\\
$N_{\rm H}$           & $1.71\pm0.04$  \\
bb $kT$               & $1.58\pm0.04$ keV \\
bb $R/d_{\rm 7~kpc}$  & $2.6\pm0.2$ km \\
brems $kT$            & $8.7\pm0.4$ keV\\
$\chi^2_{\rm r}$      & 1.42 (111 dof) \\
$\chi^2_{\rm r}$ without bb  & 3.48 (113 dof) \\
\hline
Model              & Comptonized + black body\\
$N_{\rm H}$        & $0.96\pm0.05$ \\
bb $kT$            & $1.78\pm0.06$ keV\\
bb $R/d_{\rm 7~kpc}$ & $2.4\pm0.2$ km \\
Wien $kT_{\rm W}$  & $0.57\pm0.03$ keV \\
Plasma $kT_{\rm e}$& $5.4\pm1.0$ keV\\
Plasma optical     & $3.3\pm0.6$ for disk geometry \\
\hspace{2mm}depth $\tau$       & $7.4\pm1.3$ for spherical geometry \\
Comptonization     & 0.5 for disk geometry \\
\hspace{2mm}parameter $y$      & 2.3 for spherical geometry\\
$\chi^2_{\rm r}$   & 1.21 (109 dof) \\
Flux (2-10 keV)    & $6.5\times10^{-10}$~erg~s$^{-1}$cm$^{-2}$ \\
\hspace{1cm}unabsorbed    & $7.0\times10^{-10}$~erg~s$^{-1}$cm$^{-2}$ \\
Flux (0.1-200 keV) & $9.4\times10^{-10}$~erg~s$^{-1}$cm$^{-2}$ \\
\hspace{1cm}unabsorbed    & $10.4\times10^{-10}$~erg~s$^{-1}$cm$^{-2}$ \\
$\chi^2_{\rm r}$ without bb  & 1.87 (116 dof) \\
\hline
\end{tabular}
\label{tabnfifit}
\end{table}

\begin{figure}[t]
\psfig{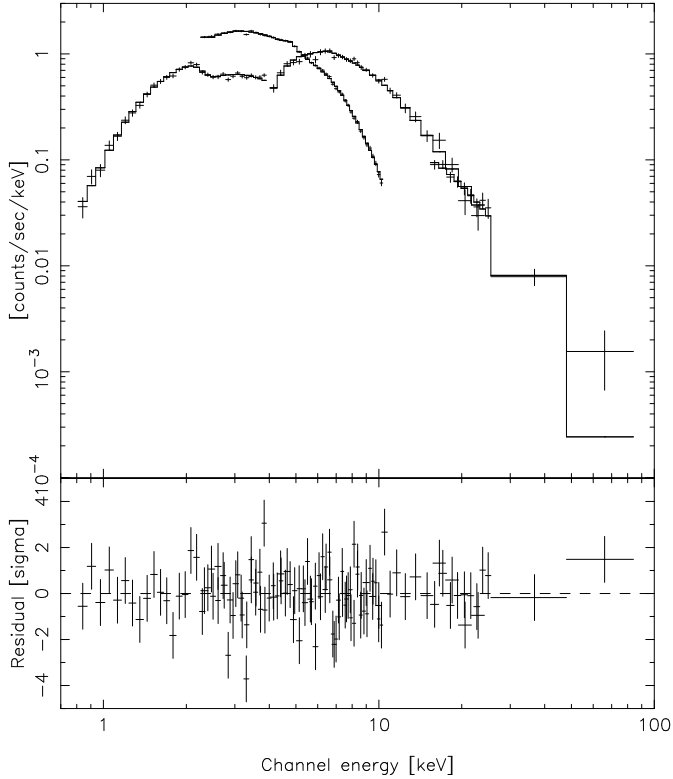}

\caption[]{Upper panel: count rate spectrum (crosses) and Comptonized spectrum
model (histogram) for the average emission at times outside the eclipse
(i.e., before 23.4 and after 32.4 ks). Lower panel: residual in units
of sigma per channel.
\label{fignfispectrum}}
\end{figure}

\subsubsection{Outside eclipse}

The spectrum outside the eclipse (Fig.~\ref{fignfispectrum}) can be
satisfactory described with various continuum models. In Table~\ref{tabnfifit},
we present the results of three models that are commonly employed in 
the study of LMXBs. 
In all models, there is the need for a broad but local component at
around 1.6-2.0 keV. We modeled this with a black body model but it can
just as well be modeled differently like for instance with a number
of narrow lines (e.g., emission lines from Si XIII and XIV)
or one very broad Gaussian line.

Because of the spectral ambiguity, we defer a scientific analysis
and merely use the parameterizations in Table~\ref{tabnfifit} to 
determine fluxes and compare the spectrum with that of other X-ray binaries.

Independent of the model, the average 0.1-200.0 keV flux is 
$9.4\times10^{-10}$~erg~s$^{-1}$cm$^{-2}$ (uncorrected for absorption). 
The column density 
$N_{\rm H}$ is between 1.0 and 1.4$\times10^{22}$~cm$^{-2}$. 
This range is consistent with the value resulting from the interstellar 
reddening to Terzan~6: for $E_{\rm B-V}=2.24$ (Barbuy et al. 
1997) with an estimated error of 0.1, $A_{\rm V}=6.94\pm0.31$ and 
$N_{\rm H}=(1.79\pm0.1)\times10^{21}A_{\rm V}=(1.2\pm0.1)\times10^{22}$~cm$^{-2}$
(according to the conversion of $A_{\rm V}$ to $N_{\rm H}$ by 
Predehl \& Schmitt 1995). The broad component at 1.6-2.0 keV, which we
modeled by black body radiation, affects the determination of $N_{\rm H}$
somewhat but we estimate that this is limited to $0.2\times10^{22}$~cm$^{-2}$.

\subsubsection{During eclipse}

Only the LECS and MECS provide data during the eclipse that are of sufficient
quality to allow a meaningful analysis. Formally,
the spectrum is consistent in shape with that outside the eclipse
($\chi^2_{\rm r}=1.18$ for 28 dof). Nevertheless, the spectrum has the
appearance of being somewhat softer. If, in the Comptonized model, 
k$T_{\rm e}$ is allowed to vary during the fit, it converges to a value of
$1.5\pm0.9$~keV ($\chi^2_{\rm r}=0.90$ for 27 dof). The F-test predicts a
probability of less than 0.02 for a chance occurrence of the improvement
in $\chi^2$. The same kind of improvement can be obtained
when leaving free the optical depth instead of the plasma temperature,
so we conclude that the nature of the softening is unclear.
The 2-10 keV flux is 
$2.3\times10^{-11}$~erg~s$^{-1}$cm$^{-2}$ (2-10 keV) or 4\% of that
outside the eclipse.

\subsubsection{During egress}

Figs.~\ref{fignfilc} and \ref{figlczoom} show that there appear to
be two different stages of egress: the fast 35~s rise and a shoulder
of a few hundred seconds. We generated separate spectra for these 
two time intervals and fitted them with the Comptonization model while 
keeping all parameters values fixed to those found for the out-of-eclipse
spectrum (Table~\ref{tabnfifit}) except $N_{\rm H}$. We used only LECS and 
MECS data because we are primarily interested in whether $N_{\rm H}$ 
changes during egress. From this we find that during the quick rise
$N_{\rm H}=(10\pm3)\times10^{22}$~cm$^{-2}$ and during the shoulder
$(1.27\pm0.07)\times10^{22}$~cm$^{-2}$. However, if we leave free
in addition the normalizations of the Comptonized spectrum, the 
sensitivity to measuring $N_{\rm H}$ is completely lost. We conclude
that we are unable to measure absorption effects during egress
in a model-independent way.

\section{Timing analysis of eclipses seen with WFC}
\label{secteclipses}

\begin{figure}[t]
\psfig{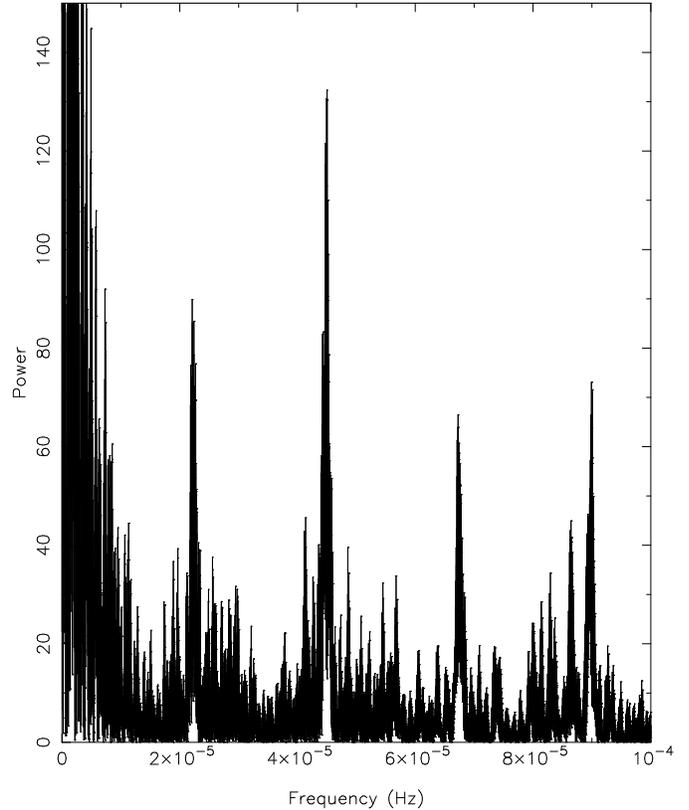}

\caption[]{Fourier power spectrum of most 2-26 keV WFC data, zoomed in at
likely orbital frequencies. The normalization is such that (white) noise 
expected from the data errors corresponds to a power of 2.
\label{figwfcpow}}
\end{figure}

\begin{figure}[t]
\psfig{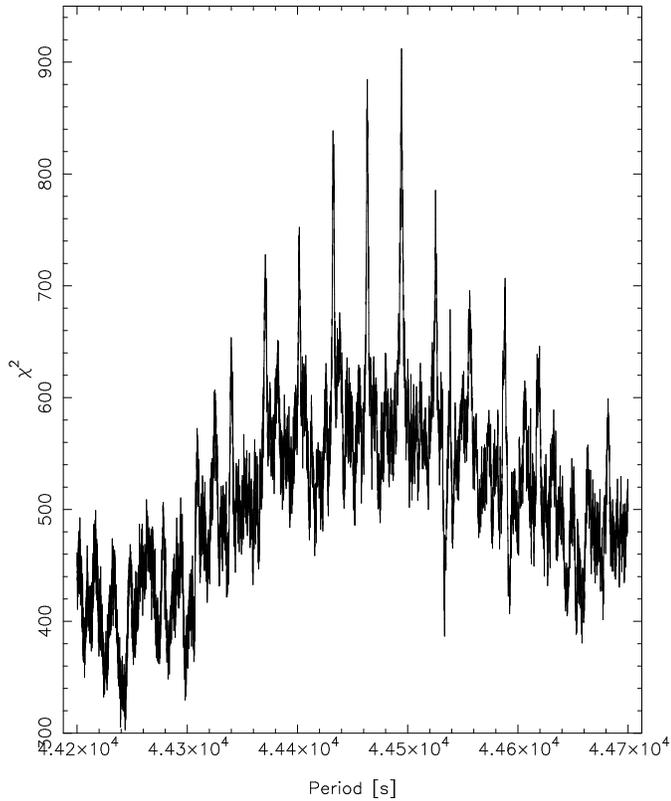}

\caption[]{Periodogram of 2-26 keV WFC data, zoomed in around most likely period
\label{figwfcperio}}
\end{figure}

\begin{figure}[t]
\psfig{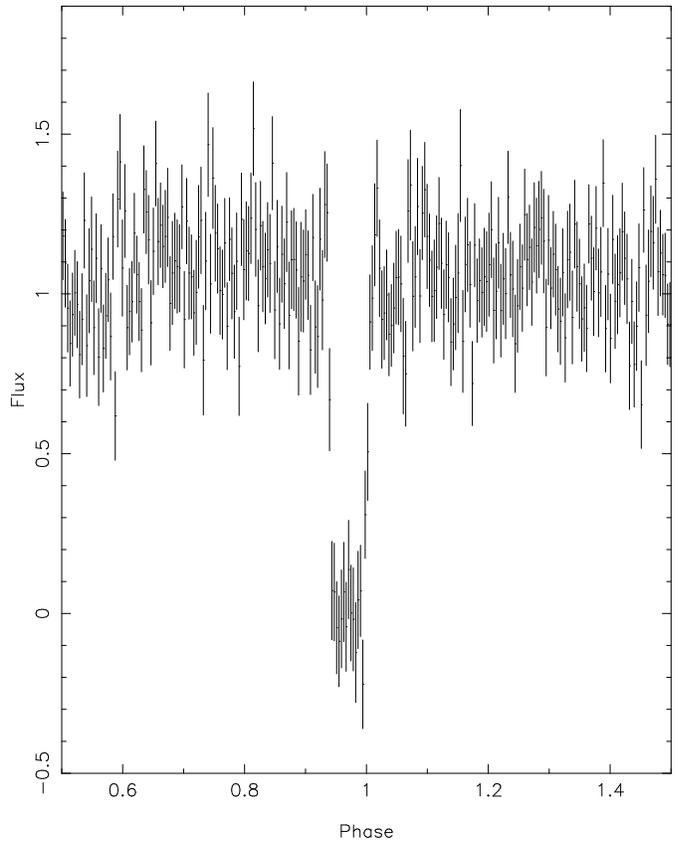}

\caption[]{Folded light curve of 2-26 WFC data, the fold period is
44494.45~s (12.360~hr). The flux has been normalized to the average.
The phase is defined with respect to the time of mid egress for
the solar system barycenter (MJD~50316.899659)
\label{figwfcfold}}
\end{figure}

The WFC observations have much longer coverage than the NFI TOO
and in principle enable the measurement of the orbital period of the binary
through the detection of more eclipses. Investigating the WFC data
at a time resolution near to the duration of the NFI-observed eclipse
is not easy for \bron\ because its peak flux is close to the 
detection limit for such a duration. In one day
of observations, the signal-to-noise ratio is at maximum about 27 so
that the 3$\sigma$ detection limit is reached when the exposure time
is of order 500~s. For a LMXB the orbital period
has a 90\% probability of being lower than 48~hr (based on the census
of White et al. 1995). 
We are fortunate with the 1996 WFC data 
because these include a continuous sequence of observations 
covering \bron\ at various off-axis angles for a duration of 9.2 days
(except for discontinuities due to earth eclipses and passes over the
South Atlantic Geomagnetic Anomaly).

We generated a light curve with a time resolution of 25~s from data
where the source was significantly detected in a complete observation 
period. This is sufficient resolution to resolve a typical eclipse.
The times were corrected to that for the solar system barycenter.
The total exposure time, corrected for times when the earth blocks the field 
of view and the SAGA is passed, is 550~ks. Naturally, the 25~s data points 
are not significant. However, if they are folded with a period of a day and 
with a phase resolution of 1/256, each light curve point represent 1.7~ks
worth of data and is expected to be be about 6 sigma. The total
time span of the light curve is 752~d (or 2.06~yr).

Fig.~\ref{figwfcpow} presents the Fourier power spectrum of the light 
curve between 0 and 10$^{-4}$~Hz. There is a strong signal at a frequency 
of about $2.25\times10^{-5}$~Hz which has a number of higher harmonics. 
Surprisingly, the harmonic at $4.5\times10^{-5}$~Hz is stronger and 
more narrowly peaked. 
We evaluated folded light curves for 1.1, 2.25 and 4.5$\times10^{-5}$~Hz,
and find that the one for 2.25$\times10^{-5}$~Hz is the one with the
deepest single eclipse. The folded light curve for 4.5$\times10^{-5}$~Hz
results in a single eclipse which is only half as deep as the one at
2.25$\times10^{-5}$~Hz and the equivalent period is, therefore, discarded 
as a possible orbital period.
To determine the period more accurately, we applied a folding technique
to search for the period with a resolution of 0.05~s. The periodogram is
given in Fig.~\ref{figwfcperio}. This shows a broad peak with narrow
spikes on top of that. The narrow spikes are uniformly spaced. The 
spacing is determined by the number of periods that are covered 
by the time span. It represents the uncertainty of the synchronization
after large data gaps. We regard this as the principal uncertainty in 
the period
determination. The two largest spikes are approximately equal in
value and are at periods of 44463.4 and 44494.45~s. From this we
determine the period to be one of these two values, with the larger
one being the most likely period because it gives the deepest eclipse.
In conclusion, the period is determined to be: $P=44494.45\pm31.05$~s
or $12.360\pm0.009$~hr. The time of mid egress interpolated from the 
WFC results to closest to the NFI observation is MJD~51062.593682.
This is only 235~s from the actual measured epoch of MJD~51062.590967. 
Thus, the WFC and the NFI data are consistent in this respect.

Fig.~\ref{figwfcfold} shows the folded light curve at $P=44494.45$~s.
From this we determine the eclipse duration to be $2.6\pm0.2$~ks.
The WFC data covers (part of) 26 eclipses over 4 outbursts.

\begin{figure}[t]
\psfig{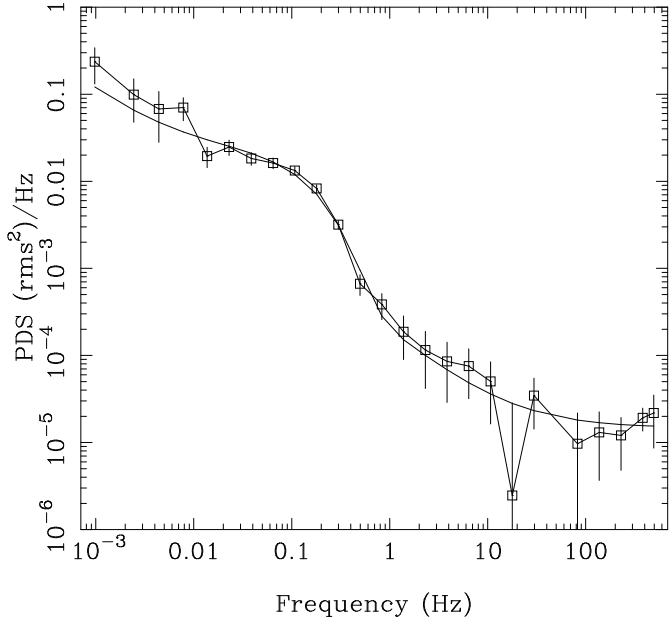}

\caption[]{Fourier power density spectrum of the intensity time series as 
measured with
{\em RossiXTE\/}-PCA in the full bandpass (2 and 60~keV). Also shown is a
model to parameterize the spectrum (see text)
\label{figpcatoo}}
\end{figure}

\section{Variability measurements with {\em RossiXTE\/}-PCA}
\label{secvar}

{\em RossiXTE\/} TOO pointed observations of \bron\ were carried out on two
separate occasions, the first on September 13, 1998, from 08:16:17 to 
09:11:00 UTC, the second on May 12, 1999, from 22:08:12 to 23:05:58 UTC.
The flux levels during both of these observations were comparable. 
We obtained high time resolution PCA data in order to search for periodic
pulsations as well as investigate the broadband X-ray variability of the
source in an attempt to constrain the nature of the compact object. 

From the two pointed observations we obtained about 6.4 ks of useful data.
We calculated fast-Fourier-transform power spectra in order to search for 
pulsations in the 
10 - 1200 Hz range. We find no significant periods in this range with an 
upper limit on any pulsed amplitude of $\approx$ 1 \% (rms). We combined data 
from both intervals to investigate the broadband variability. 
Fig.~\ref{figpcatoo} shows 
the resulting power spectrum normalized in units of (rms)$^2$ / Hz, after an
estimate of the Poisson noise level has been subtracted. We first fit a 
simple power law model including a constant offset to account for incomplete
removal of the Poisson level. The power law model alone did not provide an
adequate description of the power spectrum, mostly due to a broad excess of
power between 0.02 and 1 Hz. We modeled this additional component with an
exponential term. The resulting fit is acceptable, with
$\chi^2 = 19.4$ for 19 degrees of freedom. The amplitudes (rms) in the
power law and exponential components, integrated from $10^{-3}$ to 100 Hz, are
$4.4 \pm 0.3 \%$ and $5.7 \pm 0.4 \%$, respectively. We also searched for
kHz quasi-periodic oscillations (QPO) but found no significant features in the 
200 - 1200 Hz range. 

The X-ray variability revealed by \bron\ is qualitatively very similar to that
seen in other X-ray binaries (see for example Wijnands \& van der Klis 1999),
showing a broad excess (sometimes clearly a QPO other times properly called a
'bump') superposed on a broad band power law component. Unfortunately, this
behavior is exhibited by both black hole as well as neutron star systems, so
it is not possible to distinguish the nature of the compact source based on
the broad band X-ray variability alone. Longer observations will be required to
make more sensitive measurements to search for neutron star signatures such as
kHz QPO.

We note that no eclipse was seen with the PCA, either during the TOO
discussed in this section, or during the monitoring observations
(see Fig.~\ref{figpcalc}).

\section{Discussion} 
\label{secdis}

Despite the lack of an optical identification,
\bron\ is very likely a transient LMXB because
it is located in a globular cluster. This is also supported by
the 12.360~hr orbital period.
The eclipse provides us with further constraints on the binary orbit.
If one assumes that 1) the eclipse is caused by only the companion star;
2) the mass of the companion star is less than 
0.8~M$_\odot$ (as is expected for a main-sequence or (sub)giant star 
in a globular cluster);
3) the mass of the compact object is larger than 1.4~M$_\odot$ 
(i.e., it must be heavier than or equal to that of a neutron star); and
4) Roche geometry applies to the binary, then the calculations by Horne 
(1985) imply that the inclination angle is larger than 74\degr.

The egress out of eclipse to $\sim$90\% of the out-of-eclipse level is 
35~s. This duration is determined by the size of the emission region 
being eclipsed and the sharpness of the edge of the eclipsing object. 
Enhancements in $N_{\rm H}$ during egress would point to absorption effects 
in the atmosphere of the companion star. Unfortunately, we are not able to 
measure that. We can only determine upper limits
to the transparent part of the companion's atmosphere and the size of
the emission region. The upper limit on the relative thickness of the 
transparent part of the atmosphere is 2.7\% of the stellar radius.
The emission region should be smaller than $2\times10^3$~km
(as derived if one applies Kepler's law under the assumption that the 
combined mass of both binary components is 2~M$_\odot$).

By analogy to other eclipsing LMXBs (e.g. EXO~0748--676, Parmar
et al. 1986), it is a good hypothesis to attribute the 4\% residual
emission during the eclipse to photons scattered into the line of sight
by an accretion disk corona (ADC), plus perhaps a contribution of
interstellar dust grains below 2~keV. The emission being eclipsed would
then result from a much more confined region, very likely the inner
accretion disk. The flux contribution of the
ADC outside the eclipse would be less than 8\%. Any modulation
introduced by the partial obscuration of the ADC outside the eclipse
may, therefore, be masked by other variability of this transient source.
Possibly, the slow part of the egress as seen with the NFI may be explained
by this.
\begin{table}
\caption[]{Bursts detected from bright globular cluster sources
with the WFCs up to mid 1999\label{tabgcbursters}}
\begin{tabular}{lrl}
\hline
Name             & No.      & Exposure \\
                 & bursts   & time$^\ast$  \\
                 & detected & (Ms)     \\
\hline
Terzan 1         &  0       & 2.6 \\
Terzan 2         & 14       & 2.5 \\
Terzan 5         &  0       & 2.6 \\
Terzan 6         &  0       & 2.6 \\
Liller 1         & 27       & 2.5 \\
NGC 1851         &  0       & 1.6 \\
NGC 6440         &  3       & 2.5 \\
NGC 6441         &  0       & 2.5 \\
NGC 6624         & 24       & 2.5 \\
NGC 6652         &  3       & 2.4 \\
NGC 6712         &  0       & 1.2 \\
NGC 7078         &  0       & 0.6 \\
\hline
\end{tabular}

\noindent
$^\ast$For transients such as Terzan~6 and NGC~6440, the exposure time
includes times when the source was not active. Terzan~6 was active during
0.55~Ms.
\end{table}

If compared to other LMXBs with 
relatively faint peak fluxes, the NFI spectrum appears rather soft though not
exceptionally so. In terms of $kT_{\rm e}$ as measured with the same 
NFI instrumentation, it is for instance 
softer than SAX~J1748.9-2021 (15.5~keV, In 't Zand et al. 1999a), 
1E~1724-307 (27~keV, Guainazzi et al. 1998), GS~1826-238 (14.7~keV, In 't 
Zand et al. 1999b). On the other hand, 4U~1820-30 has a softer spectrum with
$kT_{\rm e}=2.8$~keV (Guainazzi 1999). All these sources have 
periods during which they burst, though possibly not always during the times 
when NFI spectra was taken.

One wonders why this bright globular cluster X-ray source remains,
out of 12 cases, the only one without X-ray bursts. This relates
to the question whether the compact object is a black hole candidate 
or a neutron star. If bursts would have been detected, that would have
identified for certain a neutron star. We believe the lack of bursts
is a matter of coincidence and does not help to make a statement
about the compact object. To prove this point we tabulate 
in Table~\ref{tabgcbursters} the 12 globular clusters that are known
to harbor bright X-ray sources, and the number of bursts that were
detected from each one with the WFCs. It is remarkable that, despite the 
extensive nature of the monitoring campaign of the WFCs and the
first-time detection of bursts from two of these, NGC~6652
(In 't Zand et al. 1998) and NGC 6440 (In 't Zand et al. 1999a),
no bursts were detected from 4 established globular-cluster bursters
in the Galactic center field and 3 outside that field. Apparently, 
chances are still high that bursts are missed from established 
bursters. We argue that this may very well also apply to Terzan 6.

The five likely outbursts reported in the present paper have wait times of
381, 172, 178, and 256 days. These numbers are uncertain
because, except for the last outburst, our data do not have enough
coverage to accurately determine the time of the onsets and to bridge
large data gaps due to visibility constraints for WFC observations.
However, one can conclude that there 
is a hint of a quasi periodicity of 0.5 to 0.7 years. This is a 
value not unlike that seen in other LMXBs. Future continued
bulge scan observations with the PCA will take away data gaps and
enable a better measurement of recurrence time, peak flux and
decay time and correlations between these.

The shape of the light curve of the last outburst has a fast rise and 
exponential decay. This class of outburst profiles is quite ordinary
among bright X-ray novae, accounting for about 30\% of all cases
(see review by Chen et al. 1997). The decay time constant of
18~d is common as well. The only difference is the peak flux which
is more than an order of magnitude below that of bright X-ray novae. 
The low peak flux does not seem to be chance coincidence (i.e., some bright 
recurrent X-ray novae have low-luminosity outbursts as well), its 
magnitude has been similar on five or six different occasions. The
highest peak flux was measured with the PCA. If one assumes that the
NFI-measured spectrum applies then (this is not unreasonable since
the NFI observation was close to the peak flux in another outburst),
the unabsorbed 0.1-200~keV peak luminosity is $7\times10^{36}$~\lum\
(for a distance of 7~kpc).

Despite the high inclination angle, we failed to detect dipping activity.
However, the lack of dips could be due to the insufficient sensitivity
and coverage of the data.

\section{Summary and conclusions}

We found \bron\ to be an eclipsing LMXB with an orbital period of 
12.360$\pm0.009$ hr. The eclipse duration is 0.72$\pm$0.06~hr. 
This translates to a lower limit on the inclination
angle of 74\degr, for a compact object mass of at least 1.4~M$_\odot$. 
Furthermore, \bron\ is a transient source with an unabsorbed 0.1-200~keV
peak luminosity of $7\times10^{36}$~\lum\ and a relatively short
recurrence time of half a year. The short recurrence time provides 
ample opportunity to study the source in the future. It is not yet clear
what nature the compact object is. 

Further longer observations are planned of
this system in future outbursts to 1) search for X-ray bursts 
to be able to establish the compact object as a neutron star (the bursts
may be weak if only the scattered emission is visible to us); 2) diagnose 
further the residual emission during the eclipse; 3) accurately 
determine the orbital period through more detailed eclipse timing, which may 
also enable studies of the orbital period evolution; 4) measure the spectral 
evolution during an entire outburst to find clues regarding to the faint 
nature of the source; and 5) search for dips to constrain the inclination 
angle further.

\begin{acknowledgements}
Wouter Hartmann is thanked for help with the software. This research has 
made use of SAXDAS linearized and cleaned event files (Rev.~1.1) produced
at the BeppoSAX Science Data Center. {\em BeppoSAX\/} is a joint Italian
and Dutch program. 

\end{acknowledgements}

\end{document}